\documentclass[12pt]{article}

\usepackage{amsmath}
\usepackage{amsthm}
\usepackage{amstext}
\usepackage{amsfonts}
\usepackage{amssymb}
\usepackage{afterpage}
\usepackage{rotating}
\usepackage{lscape}
\usepackage{capt-of}
\usepackage{breqn}
\usepackage{mathtools} 
\usepackage{bbm}
\usepackage{comment}
\usepackage{graphicx, color}
\usepackage{color, colortbl}
\usepackage{fancyhdr}
\usepackage[breaklinks, pdfstartview= FitH, colorlinks=true, linkcolor=blue, citecolor=blue, urlcolor=black]{hyperref}
\usepackage{tikz}

\usepackage{textcomp}
\usepackage{eurosym}

\usepackage{booktabs}
\usepackage{multirow}
\usepackage{float}

\usepackage[section]{placeins} 

\usepackage{enumerate}
\usepackage{natbib}
\newcommand{\blind}{0}

\begin{document}

\def\spacingset#1{\renewcommand{\baselinestretch}%
{#1}\small\normalsize} \spacingset{1}

\newtheorem{Theorem}{Theorem}
\newtheorem{Proposition}{Proposition}
\newtheorem{Proof}{Proof}
\newtheorem{Remark}{Remark}
\newtheorem{Assumption}{Assumption}
\newtheorem{lemma}{Lemma}
\newtheorem{Definition}{Definition}

\DeclarePairedDelimiter\ceil{\lceil}{\rceil}
\DeclarePairedDelimiter\floor{\lfloor}{\rfloor}


\if0\blind
{
  \title{\bf Testing for a Forecast Accuracy Breakdown under Long Memory}
  \author{Jannik Kreye\thanks{
    The authors gratefully acknowledge support from Deutsche Forschungsgemeinschaft (DFG, German Research Foundation) – Projektnummern 258395632 and INST 187/742-1 FUGG. We would like to thank the participants of the Statistical Week 2023 in Dortmund and the participants of the Annual Conference of the International Association for Applied Econometrics 2024 in Thessaloniki.}\hspace{.2cm}\\
    Institute of Statistics, Leibniz University Hannover\\
    and \\
    Philipp Sibbertsen \\
    Institute of Statistics, Leibniz University Hannover}
  \maketitle
} \fi

\if1\blind
{
  \bigskip
  \bigskip
  \bigskip
  \begin{center}
    {\LARGE\bf Testing for a Forecast Accuracy Breakdown under Long Memory}
\end{center}
  \medskip
} \fi

\bigskip
\begin{abstract}
We propose a test to detect a forecast accuracy breakdown in a long memory time series and provide theoretical and simulation evidence on the memory transfer from the time series to the forecast residuals. The proposed method uses a double sup-Wald test against the alternative of a structural break in the mean of an out-of-sample loss series. To address the problem of estimating the long-run variance under long memory, a robust estimator is applied. The corresponding breakpoint results from a long memory robust CUSUM test. The finite sample size and power properties of the test are derived in a Monte Carlo simulation. A monotonic power function is obtained for the fixed forecasting scheme. In our practical application, we find that the global energy crisis that began in 2021 led to a forecast break in European electricity prices, while the results for the U.S. are mixed.
\end{abstract}

\noindent%
\vfill

\newpage
\spacingset{1.8} 
\section{Introduction}

The presence of low frequency contaminations, such as structural breaks, poses a challenge in distinguishing true long memory from spurious long memory. This is due to the fact that both concepts share similar characteristics, such as a singularity in the periodogram near the zero Fourier frequency and significant autocorrelations at large lags (\cite{diebold2001long}, \cite{granger2004occasional}, \cite{mikosch2004nonstationarities}).
Tests for structural breaks are closely related to tests for forecast breakdowns, which test for structural breaks in a forecast error loss function. \cite{perron2006dealing} reviews the extensive literature on structural breaks. \cite{Perr21} extend classical tests for structural breaks to the context of forecast failure. \cite{giacomini2009detecting} propose a test to retrospectively assess whether a given forecast model provides stable forecasts by comparing in-sample and out-of-sample averages of a forecast error loss series. A forecast breakdown is defined as a decline in the out-of-sample performance of a forecasting model relative to its in-sample performance. To measure the forecast performance, the popular squared error loss function is used in this paper. Unlike structural break tests applied to the forecasting model, forecast accuracy breakdown tests allow for misspecified models and they can detect variance changes when a quadratic loss function is used. Furthermore, they assist the applied econometrician in determining the accuracy of a given forecast model and whether it requires modification. It is important to note, however, that the detection of a forecast breakdown does not necessarily indicate that the forecast model needs to be altered. For instance, \cite{christoffersen1997optimal} demonstrate that the optimality of a point forecast does not depend on the variance of the errors when a symmetric loss function is used. Nevertheless, a change in the variance can cause a forecast breakdown, since tests for forecast failures cannot distinguish between symmetric and asymmetric loss functions. \\
In practical applications, it is natural to distinguish between in-sample and out-of-sample periods. In retrospective testing, however, an artificial separation between these periods is required, since the test period serves as a pseudo-out-of-sample period. The forecasting model is estimated in the in-sample period, and to assess whether the forecast accuracy changes, it should provide stable forecasts. To obtain forecasts, \cite{Perr21} recommend using a fixed forecast scheme, since it is the only forecast scheme that leads to a monotonic power function, mainly because it ensures the maximum difference between the in-sample and out-of-sample means of the loss series. Rolling and recursive schemes can induce power losses due to contamination problems.
Under a fixed scheme, the test proposed by \cite{giacomini2009detecting} is a Wald test for a single mean change in the total loss series at a predetermined date. The total loss series combines in-sample and out-of-sample losses. The test suffers from power losses if the break date is not specified precisely. To address this issue, \cite{rossi2012out} propose a sup-Wald test that maximizes the \cite{giacomini2009detecting} test over a predefined range of potential break dates. When multiple changes occur, the test still suffers from non-monotonic power functions. For this reason, \cite{Perr21} propose a double sup-Wald (DSW) test for a structural break in the mean of the out-of-sample loss series, estimating the breakpoint with the \cite{bai1998estimating} estimator. The proposed method involves performing SW tests for each choice of the in-sample period. Ultimately, the DSW test statistic value is the largest value among all in-sample period choices. In this paper, we extend the DSW test to the long memory framework by using a memory and autocorrelation consistent (MAC) estimator of the long-run variance proposed by \cite{robinson1995gaussian} and by incorporating a long memory robust estimator of the breakpoint.
A related topic is discussed by \cite{paza2023optimal}. They obtain optimal forecasts in a long memory time series setting under discrete structural breaks. To the best of our knowledge, our study is the first to connect long memory time series with forecast failures.\\
\cite{Krus18} derive theoretical evidence for a long memory transfer from a time series to forecasts and to forecast differentials, especially in the presence of biased forecasts. We show that this transfer also applies to the squared forecast error, which is a component of a forecast differential. We consider biased and unbiased forecasts, as well as the presence and absence of a fractional cointegration relationship between a long memory time series and its forecast.\\
The remainder of the paper is organized as follows. Section \ref{LMloss} contains the theoretical derivation of the long memory transfer from the time series to the forecasts and to the squared forecast error. This justifies the need for a long memory robust forecast accuracy breakdown test, which is presented in Section \ref{test}. To derive the finite sample size and power properties of the test, we perform a Monte Carlo study in Section \ref{monte}. Section \ref{app} applies the proposed test to European and U.S. electricity prices, while Section \ref{conclusion} contains concluding remarks. All proofs are consolidated in the Appendix.

\section{Long Memory in Forecast Residuals}
\label{LMloss}
Before presenting the test to detect a change in forecast accuracy, we analyze the transfer of long memory from a time series and its forecast to the loss function under specific settings. We distinguish between biased and unbiased forecasts. Our analysis follows \cite{Krus18}, who derive long memory transfer properties for forecast error loss differentials. The proofs of all propositions are provided in the Appendix. To create a framework, we restrict the time series and its forecast to be stationary long memory processes.
\begin{Assumption}
The time series $y_t$ and the forecast $\hat{y}_t$, with $t = 1, \ldots, T$ and expectations $\mathbb{E}(y_t) = \mu_y$ and $\mathbb{E}(\hat{y}_t) = \mu_{\hat{y}}$, are causal Gaussian long memory processes having the spectral density $f_a(\lambda) \sim L_f(\lambda)|\lambda|^{-2d_a}$, for $a_t \in \{y_t, \hat{y}_t\}$, as $\lambda \rightarrow 0$, where $L_f(\lambda) \geq 0$ is a symmetric function that varies slowly at the origin. We write $a_t \sim LM(d_a)$, where $d_a \in [0; 1/2)$.
\label{longmemo}
\end{Assumption}
Assumption \ref{longmemo} is convenient since \cite{dittmann2002properties} derive long memory properties for squares and cross products of Gaussian long memory processes.
We now distinguish between the presence of fractional cointegration between the time series and its forecast and the absence of this property. A fractional cointegration relationship refers to the existence of a linear combination of long memory time series that is integrated of reduced order. 
The memory transfer results are based on the asymptotic behavior of autocovariance functions of products and squares of long memory time series. In addition, Proposition 3 of \cite{chambers1998long} can be used, as it states that the memory order of a linear combination of fractionally integrated series is equal to the maximum order of the individual components. \cite{leschinski2017memory} generalizes this result for the broader class of long memory processes and discusses the memory of products of long memory time series. We derive the long memory transfer for the squared error loss function, which we also use for our test in Section \ref{test},
\begin{equation}
    L_t = (y_t - \hat{y}_t)^2.
    \label{loss}
\end{equation}
First, consider the case of the absence of fractional cointegration.
\begin{Assumption}
If $y_t, \hat{y}_t \sim LM(d)$, then $y_t - \phi_0 - \phi_1 \hat{y}_t \sim LM(d)\ \forall\ \phi_0, \phi_1 \in \mathbb{R}$.
\label{AbsenceFCI}
\end{Assumption}
Under Assumption \ref{AbsenceFCI}, the long memory transfer is summarized in the following proposition.
\begin{Proposition}
\label{propAFCI}
Under Assumptions \ref{longmemo} and \ref{AbsenceFCI}, the squared forecast error in Eq. (\ref{loss}) is $L_t \sim LM(d_L)$, where \\
\[
d_L = \left\{\begin{array}{lr}
        \text{max}\{d_y, d_{\hat{y}}\}, & \text{if}\ \mu_{y} \neq \mu_{\hat{y}} \\
        \text{max}\{2 \text{max}\{d_y,d_{\hat{y}}\} - 1/2,\ 0 \}, & \text{if}\ \mu_{y} = \mu_{\hat{y}}.
        \end{array}\right.
\]
\end{Proposition}
\textbf{Proof.} \textit{See the Appendix.}\\
The result is obtained by replacing $y_t$ and $\hat{y}_t$ by their centered series $a_t^* = a_t - \mu_a$ for $a_t \in \{y_t,\hat{y}_t\}$.
As shown in the Appendix, the squared forecast error in Eq. (\ref{loss}) can then be rearranged as
\begin{equation}
    L_t= 2[y_t^*(\mu_y-\mu_{\hat{y}}) - \hat{y}_t^*(\mu_y-\mu_{\hat{y}})] - 2y_t^*\hat{y}_t^* + y_t^{*^2} + \hat{y}_t^{*^2}  +\ \text{const}.
    \label{sqrterrorAFCI}
\end{equation}
Proposition \ref{propAFCI} shows that if the forecast is biased, the memory order of the original terms dominates.
If the forecast is unbiased, the first part of Eq. (\ref{sqrterrorAFCI}) drops and the product and squared series remain. As shown in the Appendix, the memory orders of the squared series dominate the memory of the product series and the result follows.\\
We now relax Assumption \ref{AbsenceFCI} and allow for fractional cointegration between the series and the forecast.
\begin{Assumption}
Let $x_t$ be a stationary long memory process according to Assumption \ref{longmemo} with memory parameter $d_x$. Then, if $y_t$ and $\hat{y}_t$ are $LM(d_x)$ and fractionally cointegrated, they can be represented as $y_t = \beta_y + \kappa_y x_t + \epsilon_y$ and $\hat{y}_t = \beta_{\hat{y}} + \kappa_{\hat{y}} x_t + \epsilon_{\hat{y}}$ with $\kappa_y,\kappa_{\hat{y}} \neq 0$ and $\epsilon_y, \epsilon_{\hat{y}}$ are centered causal Gaussian long memory processes with $1/2 > d_x > d_{\epsilon_y}, d_{\epsilon_{\hat{y}}}$. We write $y_t,\hat{y}_t \sim FCI(d_x, d_x - \delta)$.
\label{assumpFCI}
\end{Assumption}
$d_x - \delta$ refers to the memory order of the linear combination of $y_t$ and $\hat{y}_t$, and they are fractionally cointegrated for $\delta > 0$. 
As can be seen from Assumption \ref{assumpFCI}, $x_t$ is the common factor that drives the long memory properties of $y_t$ and $\hat{y}_t$. We restrict the fractional cointegration to a form where the two series can be represented as linear functions of their common factor. In fact, our previous Assumption \ref{AbsenceFCI} of the absence of fractional cointegration may be inappropriate in a forecasting setup. It follows directly from Assumption \ref{assumpFCI} that
\begin{equation*}
    y_t - \hat{y}_t = \beta_y - \beta_{\hat{y}} + x_t(\kappa_y - \kappa_{\hat{y}}) + \epsilon_y - \epsilon_{\hat{y}}.
\end{equation*}
There are three cases to consider under fractional cointegration. 
\begin{enumerate}[(i)]
  \item Biased forecast and $\kappa_y \neq \kappa_{\hat{y}}$,
  \item biased forecast and $\kappa_y = \kappa_{\hat{y}}$,
  \item unbiased forecast and $\kappa_y = \kappa_{\hat{y}}$.
\end{enumerate}
The memory transfer results for the three cases are obtained by substituting the relations from Assumption \ref{assumpFCI} into the squared forecast error in Eq. (\ref{loss}). By rearranging the resulting equation, we obtain an expression similar to that in Eq. (\ref{sqrterrorAFCI}). We obtain the following result.
\begin{Proposition}
Under Assumptions \ref{longmemo} and \ref{assumpFCI}, the squared forecast error in Eq. (\ref{loss}) is $L_t \sim LM(d_L)$, where
\[
    d_L = \left\{\begin{array}{lr}
    d_x, & \text{if}\ \mu_y \neq \mu_{\hat{y}}\ \text{and}\ \kappa_{y} \neq \kappa_{\hat{y}}\\
    \tilde{d}, & \text{if}\ \mu_y \neq \mu_{\hat{y}}\ \text{and}\ \kappa_y = \kappa_{\hat{y}}\\
    \text{max}\{2d_x-1/2,0\}, & \text{if}\ \mu_y = \mu_{\hat{y}}\ \text{and}\ \kappa_y = \kappa_{\hat{y}}
    \end{array}\right.
\]
for some $0 \leq \tilde{d} < d_x$.
\label{propFCI}
\end{Proposition}
\textbf{Proof.} \textit{See the Appendix.}\\
In case of a biased forecast with $\kappa_y \neq \kappa_{\hat{y}}$, the memory order of $x_t$ dominates.
In the other two cases, the memory is reduced. Under a biased forecast and $\kappa_y = \kappa_{\hat{y}}$, the resulting memory parameter of the squared forecast error is $\tilde{d}$. The exact order of $\tilde{d}$ cannot be determined because squares and products of the errors $\epsilon_y$ and $\epsilon_{\hat{y}}$ are involved, for which the memory orders are unknown. In the remaining case of an unbiased forecast with $\kappa_y = \kappa_{\hat{y}}$, the forecast is an accurate prediction of the underlying series. Here, the memory of $x_t^{*^2}$ dominates, but this memory is also smaller than that of $x_t^*$, as shown in the Appendix.\\
In summary, our main results in Propositions \ref{propAFCI} and \ref{propFCI} show that long memory can be transferred from the underlying time series to the forecast and finally, to the squared forecast error. The memory transfer depends crucially on the (un)biasedness of the forecast and the presence of fractional cointegration.

\section{Double Sup-Wald Test}
\label{test}
After deriving the long memory transfer from the time series to the forecast and finally to the squared forecast error, we now turn to the detection of a forecast failure.
The accuracy of the point forecasts is evaluated in the out-of-sample period. We use the popular out-of-sample squared error loss function 
\begin{equation}
\label{olosses}
    L_{t+\tau}^o (m)= (y_{t+\tau} - \hat y_{t+\tau})^2,
\end{equation}
where $L^o (m) = L_{m+\tau}^o,\ldots,L_{T}^o$ is the out-of-sample loss sequence of size $n = T-m-\tau+1$. Under the null hypothesis,
\[
H_0: E[L_t^o] = \mu_0, \qquad \forall t = \tau + 1, \ldots, T-\tau +1,
\]
there is no structural break in the mean of the out-of-sample loss series.
The alternative hypothesis is given by:
\[
H_1: E[L_t^o] \neq E[L_{t+1}^o]
\]
for at least one $t = \tau + 1, \ldots, T-\tau$. Thus, our test is capable of detecting a single break in the forecast performance, but it does not rule out the existence of multiple breaks. We do not take into account the presence of multiple breaks because there is no long memory robust breakpoint estimator yet.\\
The underlying test for a break in the predictive accuracy is a Wald-type test proposed by \cite{Perr21}.
To compute the test statistic, the out-of-sample loss series is calculated for each in-sample period of size $m$ in an interval $[m_0,m_1]$, where the range between $m_0$ and $m_1$ is usually a fraction of the size of the largest out-of-sample period. The SW test is then applied to each out-of-sample loss series, and we take the maximum over all SW test statistics for all choices of the in-sample period, resulting in a DSW test. 
The test statistic is given by
\begin{equation*}
DSW_{L^o(m)} = \max_{m \in [m_0, m_1]} SW_{L^o(m)},
\end{equation*}
where
\begin{equation}
SW_{L^o(m)} = \max_{T_b(m) \in [m+\tau + \varepsilon n, m+\tau +(1-\varepsilon)n]} \frac{SSR_{L^o(m)}-SSR(T_b(m))_{L^o(m)}}{\hat{V}_{L^o(m)}}.
\label{supWald}
\end{equation}
$\varepsilon$ is a trimming parameter that is set to $0.1$ throughout the simulations and application. For a given choice of $m$, the test maximizes the difference between the demeaned restricted sum of squared residuals of the out-of-sample loss series $\left(SSR_{L^o(m)}\right)$ and the demeaned unrestricted SSR of the out-of-sample loss series $\left(SSR\left(T_b(m)\right)_{L^o(m)}\right)$. $T_b(m)$ denotes the breakpoint, which results from a long memory robust CUSUM test proposed by \cite{horvath1997effect} for Gaussian processes and extended by \cite{wang2008change} to general linear processes. Both papers derive the limit distributions that converge to the supremum of a fractional Brownian bridge. The difference between the SSR is standardized by the MAC long-run variance estimate of the sample mean proposed by \cite{robinson2005robust} and \cite{abadir2009two}, which is applied to the demeaned out-of-sample loss series with a break at time $T_b(m)$. The MAC estimator is defined as
\begin{equation*}
\hat{V}{_{L^o(m)}}(d_L) = \hat{b}_w(d_L) p(d_L),
\end{equation*}
where
\begin{equation*}
\hat{b}_w(d_L) = w^{-1}\sum_{j=1}^w \lambda_j^{2d_L} I_n(\lambda_j) 
\end{equation*}
and 
\begin{equation*}
I_n(\lambda_j) = (2 \pi n)^{-1} \left| \sum_{t=1}^n e^{it\lambda_j} L^o_t(m) \right|^2
\end{equation*}
is the periodogram, $\lambda_j = 2\pi j/n$ are the Fourier frequencies and the bandwidth $w$ converges to infinity and satisfies $w = o(n/(log\ n)^2)$.
Finally,
\begin{equation*}
 p(d_L) = \left\{ 
\begin{array}{lr}
2 \frac{\Gamma(1-2d_L)sin(\pi d_L)}{d_L(1+2d_L)}, &\ \text{if}\ d_L \neq 0, \\
2 \pi, &\ \text{if}\ d_L = 0.  \\
\end{array}
\right.\\   
\end{equation*}
The MAC estimator requires an estimate of the long memory parameter of the out-of-sample loss series under the alternative. For this purpose, we use the local Whittle estimator of \cite{kunsch1987statistical} and \cite{robinson1995gaussian},
\begin{equation*}
\hat{d}_L = \underset{d_L\in[-1/2,1/2]}{\text{argmin}} \log\left(\frac{1}{w_{\scriptscriptstyle LW}} \sum_{j=1}^{w_{\scriptscriptstyle LW}}j^{2d_L}I_n(\lambda_j)\right) - \frac{2d_L}{w_{\scriptscriptstyle LW}} \sum_{j=1}^{w_{\scriptscriptstyle LW}}\log j.
\end{equation*}
For the bandwidth choices, we follow the simulation results of \cite{abadir2009two} and choose $w=\floor{n^{0.8}}$ for the bandwidth of the MAC estimator and $w_{\scriptscriptstyle LW}=\floor{n^{0.65}}$ for the bandwidth of the local Whittle estimator.\\
Since we are trying to detect a change in forecast accuracy, for each $m \in [m_0,m_1]$ the selected in-sample period should be stable in order to provide stable forecasts. To obtain stable forecasts, a fixed forecasting scheme should be chosen where the in-sample period consists of observations $1,\ldots,m$, because a rolling and a recursive scheme ensure that the mean of the in-sample losses approaches the mean of the out-of-sample losses rather quickly. This prevents forecast accuracy breakdown tests from generating power under a rolling and a recursive scheme. A fixed scheme may not optimize a given loss function, but that is not the goal of forecast breakdown tests.\\
We require the following assumption and additionally that $T, m\ \text{and}\ n$ converge to infinity at the same rate.
\begin{Assumption} \label{assump}
Under the null hypothesis of no change in forecast accuracy, it holds for the total loss series $L_{t+\tau} \equiv \left\{L_{t+\tau}\right\}_{t=1}^{T-2\tau+1}$: $E\left[L_{t+\tau} \right] = \mu\ \forall t,\  T^{-1+d_L} E\left[\sum^{\floor{r\left(T-2\tau+1\right)}}_{t=1} \left(L_{t+\tau} - \mu\right)\right]^2\\ \stackrel{p}{\rightarrow} r\Omega,\ \text{as}\ T \rightarrow \infty$, for $r \in [0,1]$ with $\tau$ fixed, where $\stackrel{p}{\rightarrow}$ denotes convergence in probability, $m \in [m_0,m_1]$, $\Omega$ is a full rank nonrandom matrix and $T^{-1/2+d_L} \sum^{\floor{r\left(T-2\tau+1\right)}}_{t=1} \left(L_{t+\tau} - \mu\right) \Rightarrow \Omega^{1/2} W_{d_L}(r)$, where $\Rightarrow$ denotes weak convergence in distribution and $W_{d_L}(r)$ is a fractional Brownian motion defined on $r$ with memory parameter $d_L$.
\end{Assumption}
The limit distribution of the test is derived in the Appendix and summarized in the following proposition.
\begin{Proposition} \label{Wtest}
Under Assumption \ref{assump}, we have for the DSW test
\begin{equation*} \label{eq:limdist}
 T^{d_L} DSW \Rightarrow \sup_{\mu \in [0, \Bar{\mu}]}\sup_{\lambda \in [\mu + \varepsilon (1-\mu), 1-\varepsilon (1-\mu)]}\frac{\left[(\lambda-\mu)W_{d_L}(1) + (1 - \lambda)W_{d_L}(\mu) - (1-\mu)W_{d_L}(\lambda)\right]^2}{(1-\lambda)(1-\mu)(\lambda-\mu)},
\end{equation*}
as $T,m,n \rightarrow \infty$ at the same rate. $W_{d_L}(r)$ is a fractional Brownian motion defined on $r \in [0,1]$ and $\Bar{\mu} = \lim_{T \rightarrow \infty} \frac{m_1-m_0}{n_0}$ with $n_0 = T - m_0 - \tau + 1$.
\end{Proposition}
\textbf{Proof.} \textit{See the Appendix.}

\section{Monte Carlo Study}
\label{monte}
To examine the finite sample size and power properties of our proposed test, we conduct a Monte Carlo simulation study. 
\begin{table}[h]
\begin{center}
\begin{tabular}{@{}lcccc@{}}
\toprule
$\alpha$/d    & 0.1    & 0.2    & 0.3    & 0.4  \\ \midrule
10\% & 7.572 & 7.686 & 7.475 & 7.410     \\
5\%  & 9.022 & 9.335 & 9.095 & 9.167     \\
1\%  & 12.693 & 12.837 & 13.046 & 12.824 \\ \bottomrule
\end{tabular}
\caption{Critical values for FWN processes with $5{,}000$ repetitions, $T=1{,}000$, $m_0 = \floor{0.2T}$, $\Bar{\mu}=0.3$, $\tau=1$ and $\varepsilon=0.1$.}
\label{crit}
\end{center}
\end{table}
The critical values in Table \ref{crit} are obtained under the null hypothesis with $5{,}000$ repetitions of $T=1{,}000$ observations of fractional white noise (FWN) processes with a fractional parameter $d \in \{0.1,0.2,0.3,0.4\}$, Gaussian innovations and a significance level of $10\%$, $5\%$, and $1\%$. In addition, we account for short-run dynamics by adding an autoregressive parameter $\phi \in \{0.1,0.2,0.3,0.4\}$ to the FWN processes. The corresponding critical values can be found in Table \ref{critAR} in the Appendix. The critical values, the size and the power results are obtained with $m_0 = \floor{0.2T}$, so that $20\%$ of the sample data is used for the smallest choice of the in-sample period, $\Bar{\mu}=0.3$, which implies that $30\%$ of the largest out-of-sample period is used for the window defining $m_1$, the size of the largest in-sample period. We consider a forecast horizon of $\tau=1$ and a trimming value of $\varepsilon=0.1$. Changing these values slightly does not significantly change the results. We only report results for a fixed forecast scheme because, similar to \cite{Perr21}, we find that this is the only forecast scheme that ensures a monotonic power function. We elaborate on the loss of power under a recursive and a rolling forecast scheme in the supplementary material.\\
We conduct a local power analysis comparing the \cite{Perr21} test with the heteroscedasticity and autocorrelation consistent estimator of \cite{andrews1991heteroskedasticity} for the long-run variance correction factor in Eq. (\ref{supWald}) and the \cite{bai1998estimating} breakpoint estimator (hereafter abbreviated as HAC test) with our proposed test (hereafter abbreviated as MAC test). For the size and the power we use $2{,}000$ Monte Carlo repetitions and $T=500$ observations. Moreover, we let $d \in \{0.1,0.2,0.3,0.4\}$ and $\phi \in \{0,0.1,0.2,0.3,0.4\}$ and we consider mean shifts in the simulated autoregressive fractionally integrated moving average (ARFIMA) processes starting from the three hundredth observation of increasing size $k T^{d-1/2}$ for $k = 0,1,...,25$.\\
\begin{table}[]
\setlength\tabcolsep{0pt}
\begin{tabular*}{\linewidth}{@{\extracolsep{\fill}} @{}llcccccccc@{} }
\toprule
         & \multicolumn{9}{c}{$\phi=0$} \\
         & \multicolumn{4}{c}{MAC} & & \multicolumn{4}{c}{HAC} \\ \cmidrule{2-5} \cmidrule{7-10}
$\alpha$/d & \multicolumn{1}{c}{0.1} & 0.2 & 0.3 & 0.4 & & 0.1 & 0.2 & 0.3 & 0.4 \\ \cmidrule{1-5} \cmidrule{7-10}
10\% & 0.072 & 0.092 & 0.092 & 0.088 & & 0.111 & 0.106 & 0.112 & 0.128 \\
5\%  & 0.043 & 0.048 & 0.043 & 0.052 & & 0.062 & 0.062 & 0.065 & 0.071  \\
1\%  & 0.011 & 0.011 & 0.013 & 0.011 & & 0.013 & 0.025 & 0.020 & 0.015 \\ \cmidrule{1-5} \cmidrule{7-10}
         & \multicolumn{9}{c}{$\phi=0.1$} \\
10\% & 0.082 & 0.086 & 0.090 & 0.130 & & 0.112 & 0.114 & 0.130 & 0.174\\
5\%  & 0.046 & 0.045 & 0.049 & 0.065 & & 0.066 & 0.068 & 0.072 & 0.101 \\
1\%  & 0.012 & 0.012 & 0.012 & 0.022 & & 0.021 & 0.013 & 0.022 & 0.050 \\ \cmidrule{1-5} \cmidrule{7-10}
         & \multicolumn{9}{c}{$\phi=0.2$} \\
10\% & 0.090 & 0.084 & 0.095 & 0.119 & & 0.102 & 0.112 & 0.140 & 0.168 \\
5\%  & 0.053 & 0.053 & 0.045 & 0.071 & & 0.070 & 0.070 & 0.086 & 0.110 \\
1\%  & 0.012 & 0.015 & 0.009 & 0.018 & & 0.014 & 0.020 & 0.028 & 0.039 \\ \cmidrule{1-5} \cmidrule{7-10}
         & \multicolumn{9}{c}{$\phi=0.3$} \\
10\% & 0.079 & 0.087 & 0.096 & 0.104 & & 0.112 & 0.107 & 0.136 & 0.154 \\
5\%  & 0.056 & 0.052 & 0.044 & 0.044 & & 0.063 & 0.063 & 0.073 & 0.096 \\
1\%  & 0.012 & 0.012 & 0.012 & 0.012 & & 0.020 & 0.015 & 0.022 & 0.040 \\ \cmidrule{1-5} \cmidrule{7-10}
         & \multicolumn{9}{c}{$\phi=0.4$} \\
10\% & 0.078 & 0.084 & 0.096 & 0.096 & & 0.114 & 0.113 & 0.128 & 0.135 \\
5\%  & 0.042 & 0.041 & 0.045 & 0.045 & & 0.060 & 0.059 & 0.063 & 0.081 \\
1\%  & 0.005 & 0.009 & 0.016 & 0.012 & & 0.013 & 0.015 & 0.025 & 0.035 \\ \bottomrule
\end{tabular*}
\caption{Empirical size of the MAC and HAC test for ARFIMA($\phi$,$d$,$0$) processes with $2{,}000$ repetitions, $T=500$, $m_0 = \floor{0.2T}$, $\Bar{\mu}=0.3$, $\tau=1$ and $\varepsilon=0.1$.}
\label{size}
\end{table}
Table \ref{size} shows the empirical size for nominal significance levels of $10\%$, $5\%$, and $1\%$. On average, the empirical size of our MAC test is closer to the nominal size than the empirical size of the HAC test. At the same time, our test is slightly undersized on average, while the HAC test is oversized for all parameter combinations. The liberalness of the HAC test increases on average with increasing fractional parameter value. For increasing short-run dynamics, the liberalness of the HAC test seems to decrease, at least for $d=0.4$.\\
\begin{figure}
	    	\centering
		\input{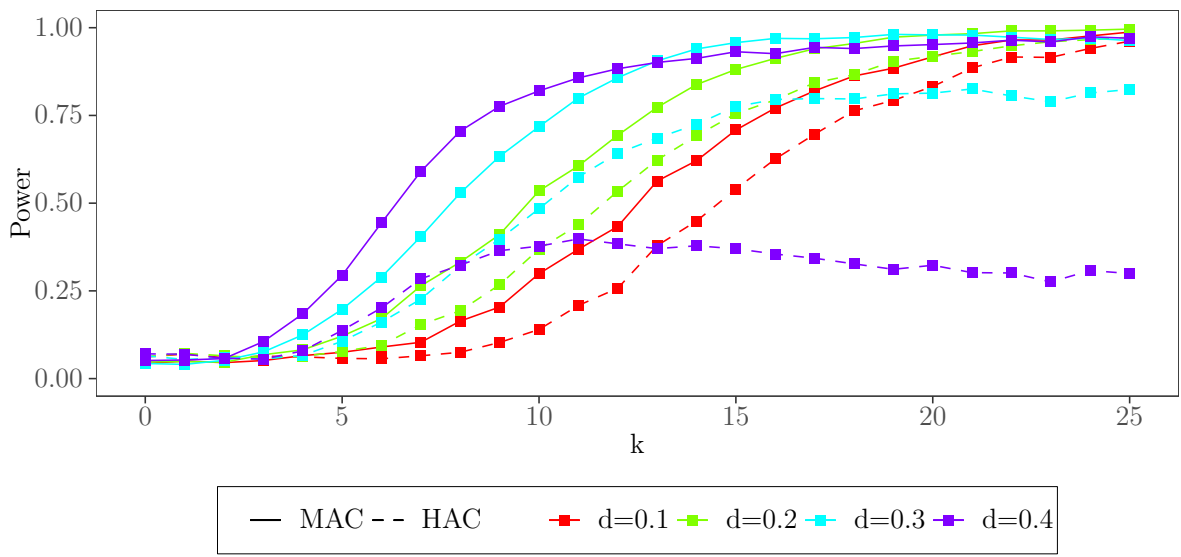}
		\caption{Local power for a mean break in FWN processes with long memory parameter $d$, $2{,}000$ repetitions, $T=500$, $m_0 = \floor{0.2T}$, $\Bar{\mu}=0.3$, $\tau=1$ and $\varepsilon=0.1$.}
		\label{FWNpower}
	\end{figure}
Figure \ref{FWNpower} shows the local power functions for FWN processes with a nominal significance level of 5\%. Overall, the MAC test is superior in terms of power for all values of the fractional parameter. The power differences increase with increasing fractional parameter value and the HAC test with $d=0.4$ even loses power.\\
 \begin{figure}
	    	\centering
		\input{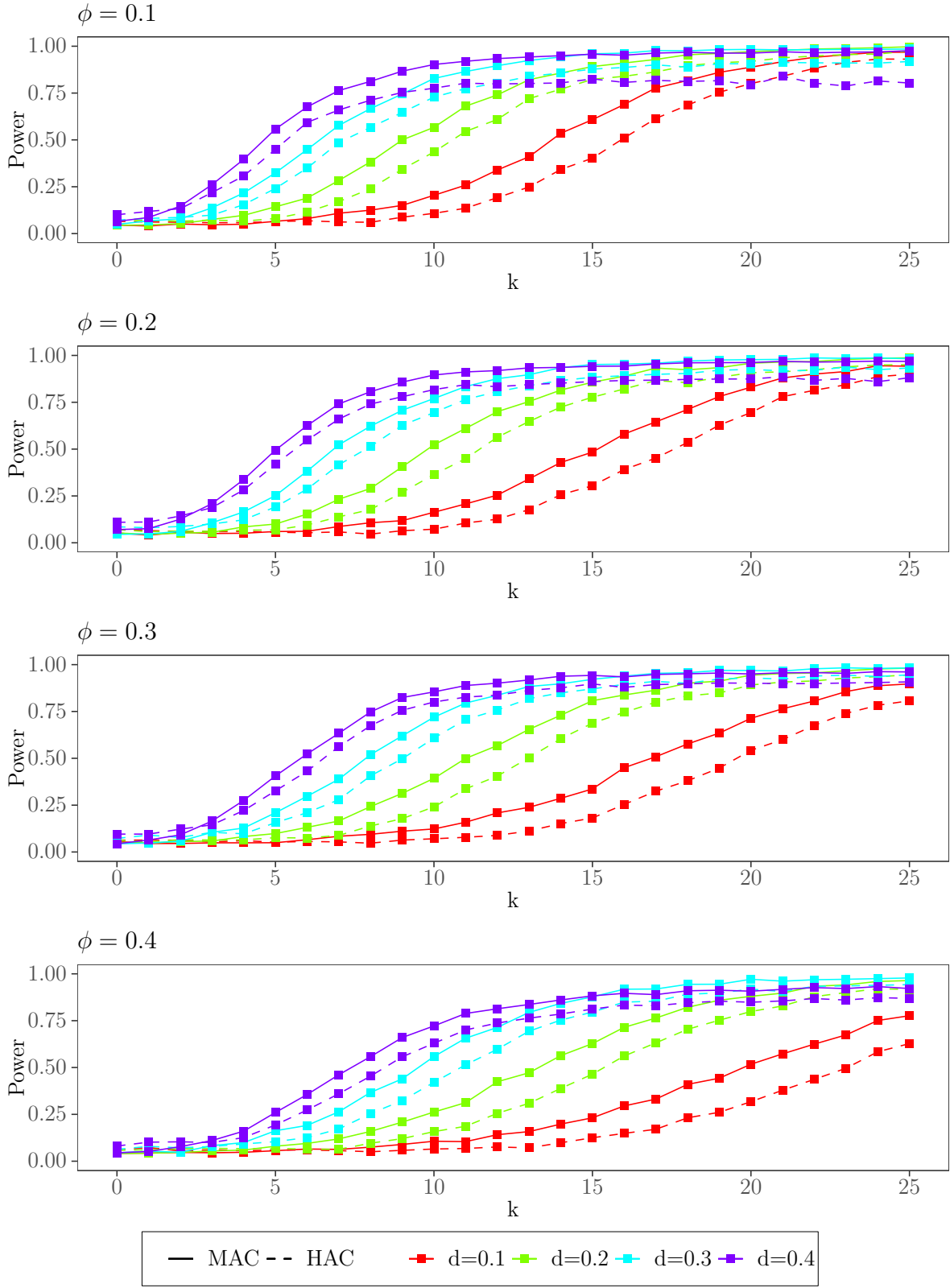}
		\caption{Local power for a mean break in ARFIMA($\phi$,$d$,$0$) processes with $2{,}000$ repetitions, $T=500$, $m_0 = \floor{0.2T}$, $\Bar{\mu}=0.3$, $\tau=1$ and $\varepsilon=0.1$.}
       \label{ARpower} 
\end{figure}
Figure \ref{ARpower} introduces short-run dynamics of increasing size to the simulated FWN processes. Our test is still superior in terms of power, but the power of the HAC test increases overall. On average, both tests lose power when the short-run dynamics increase.\\
We also considered variance changes in the simulated ARFIMA processes, as \cite{giacomini2009detecting} state that forecast accuracy breakdown tests with a squared error loss function can detect variance changes as well. However, as we have shown in Section \ref{LMloss}, a memory reduction to zero is possible when the forecast is unbiased. Both tests are still consistent under a variance break, but they perform similarly in terms of power, so we do not report the results.
\section{Breaks in Forecast Accuracy during the Global Energy Crisis}
\label{app}
The global energy crisis that commenced in 2021 resulted in a significant surge in electricity prices. In this analysis, we apply our proposed test to European and U.S. daily wholesale electricity prices with the objective of identifying differences in the severity and timing of the changes in forecasting performance in response to the crisis. The European electricity price series include data from Denmark, France, Germany, the Netherlands, and Norway. They are obtained from the German Federal Network Agency. The U.S. electricity price hubs considered are Mass Hub, Mid-C, Palo Verde, PJM West, and SP-15. The data set was obtained from the U.S. Energy Information System. The regions the price hubs cover can be found on the organization's website. It should be noted that additional U.S. price hubs are not considered in this analysis due to incomplete data. In order to facilitate comparison with the U.S. data, weekends have been removed from the European data set. In view of the sharp rise in inflation following the global pandemic, all price series were adjusted for inflation using an electricity-specific price index. The initial year of our analysis, 2015, has been designated as the base year. The electricity price indices for the European countries are obtained from the corresponding national statistical authorities\footnote{Statistics Denmark, The National Institute of Statistics and Economic Studies (France), Statistisches Bundesamt (Germany), Centraal Bureau voor de Statistiek (The Netherlands), Statistics Norway.}, while the corresponding index for the U.S. has been obtained from the U.S. Bureau of Labor Statistics. 
\begin{figure}
    \centering
    \includegraphics[width=1\textwidth]{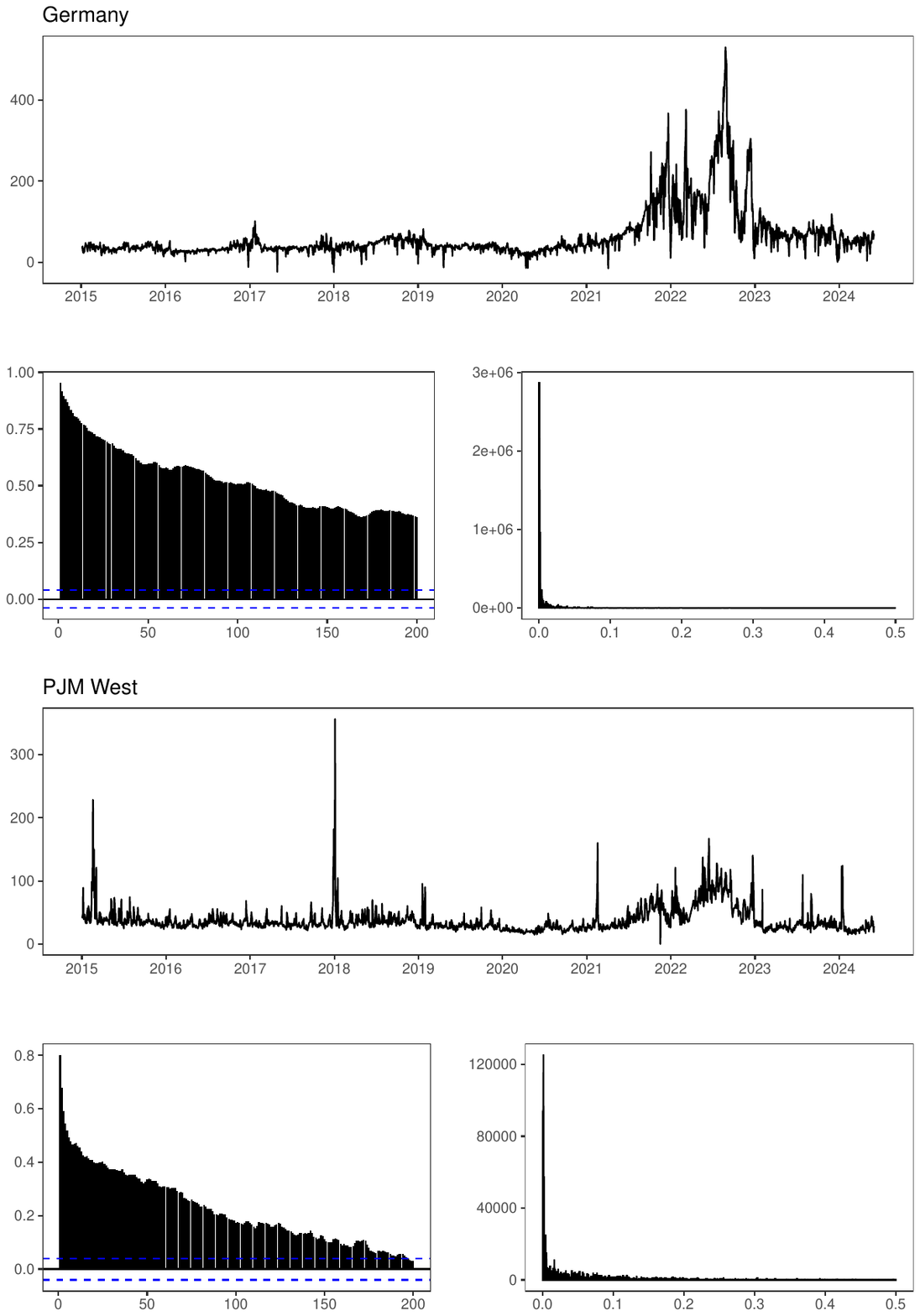}
    \vspace{-50pt}
    \caption{Price series, ACF and periodogram for Germany and PJM West.}
    \label{GerPjm}
\end{figure}
Figure \ref{GerPjm} shows the inflation-adjusted electricity price series for Germany and the price hub PJM West, spanning from 01/01/2015 to 05/31/2024. It also illustrates the autocorrelation function (ACF) and the periodogram of the corresponding series. The remaining price series, ACFs and periodograms are shown in Figures \ref{DenFra}, \ref{NetNor}, \ref{MasMid} and \ref{PalSp15} in the Appendix. 
The series are highly persistent with a pole at frequency zero, indicating that the series possess long memory properties. The characteristic weekly seasonality of electricity prices, which arises from the change in electricity demand and generation during the week as opposed to the weekend, is not evident in the data set, as weekends are not included. The seasonal component of electricity prices has already been analyzed by \cite{haldrup2006regime} in a long memory framework. \\
For the application of our test, we select the smallest in-sample period ending on 01/01/2018 and the largest in-sample period ending on 12/31/2019. The critical values for each time series are obtained by simulating $5{,}000$ ARFIMA paths with $T=1{,}000$ observations, $m_0 = \floor{0.2T}$, $\Bar{\mu}=0.3$, $\tau=1$ and $\varepsilon=0.1$.
\begin{table}[]
\setlength\tabcolsep{0pt}
\begin{tabular*}{\linewidth}{@{\extracolsep{\fill}} @{}lccccc@{} }
\toprule
Electricity Price Series  & 90\% Crit. & 95\% Crit. & 99\% Crit. & Test Statistic & Break Date \\ \midrule
Mass Hub & 11.004 & 13.747 & 21.940 & 20.852 & 12/06/2021 \\
Mid-C & 11.325 & 14.348 & 24.593 & 8.914 & - \\
Palo Verde & 10.527 & 12.841 & 18.209 & 3.901 & - \\
PJM West & 10.159 & 11.946 & 16.208 & 22.567 & 02/10/2021 \\
SP-15 & 10.324 & 12.327 & 18.910 & 10.657 & 12/08/2020 \\
Denmark & 10.778 & 13.273 & 21.481 & 262.739 & 09/22/2021\\
France & 11.964 & 14.980 & 25.609 & 91.794 & 10/05/2021\\
Germany & 9.807 & 11.828 & 16.343 & 248.031 & 09/22/2021\\
The Netherlands & 10.175 & 12.187 & 16.468 & 152.034 & 07/28/2021 \\
Norway & 12.546 & 16.145 & 34.246 & 127.330 & 11/26/2021 \\ \bottomrule
\end{tabular*}
\caption{Critical values, test statistic and corresponding break date for the U.S. and European electricity price series.}
\label{results}
\end{table}
Table \ref{results} shows the critical values, the test statistic and the break date if the test statistic is significant at any significance level. The European price series are subject to a break in forecast accuracy that is between the end of July and the end of November 2021. Thus, the breaks are before the Russian invasion of Ukraine on 02/24/2022 and the corresponding European Union sanctions against Russia. The increase in global electricity prices in 2021 is due to increased demand for energy following the pandemic, which could not be met by supply. Another factor that may have had an impact on electricity prices in the European Union is the start of Phase IV of the EU ETS in 2021. Phase IV involves a higher, linearly increasing reduction factor for the EU-wide cap on emission allowances. As a result of the cap, the price per ton of CO$_2$ and the price of gas have risen sharply, while high gas prices drive up electricity prices through the merit order effect. For the U.S. price series, a break is identified in 2021 for Mass Hub and PJM West. The break in August 2020 for SP-15 is statistically significant at the $10 \%$ level.
\section{Conclusion}
\label{conclusion}
In this paper, we propose a DSW test for detecting a forecast accuracy breakdown under a long memory time series setting. To account for the long memory properties, we standardize the test statistic with a MAC long-run variance estimate of the sample mean. The corresponding breakpoint is estimated with a long memory robust CUSUM test. The finite sample size and power properties of the test are derived in a Monte Carlo simulation. We show that our approach is superior in terms of size and power to a HAC estimate for the long-run variance if the underlying process exhibits long memory properties. The practical relevance of the method is demonstrated by applying the test to European and U.S. electricity prices. We find that European countries were more exposed to the global energy crisis that started in 2021.\\
There are ample opportunities to extend our flexible approach. The quadratic loss function may be replaced by any other loss function that the applied econometrician deems appropriate. Future research should also allow for multiple breaks, which may also occur in the in-sample period.

\section*{Appendix}

\subsection*{Proofs}

\begin{proof}[\textbf{Proof of Proposition \ref{propAFCI}}]
The centered series $a_t^* = a_t - \mu_a$ allows to reformulate the squared forecast error in Eq. (\ref{loss}).
\begin{align}
\begin{split}
\label{sqrterror}
L_t &= (y_t - \hat{y}_t)^2 \\ 
&= (y_t^\ast + \mu_y)^2 + (\hat{y}_t^\ast + \mu_{\hat{y}})^2 - 2\left[(y_t^\ast + \mu_y) (\hat{y}_t^\ast + \mu_{\hat{y}})\right] \\
&= \underbrace{2[y_t^*(\mu_y-\mu_{\hat{y}}) - \hat{y}_t^*(\mu_y-\mu_{\hat{y}})]}_{\text{I}} - \underbrace{2y_t^*\hat{y}_t^*}_{\text{II}} + \underbrace{y_t^{*^2} + \hat{y}_t^{*^2}}_{\text{III}}  +\ \text{const}.
\end{split}
\end{align}
We will now consider each component (I, II and III) of Eq. (\ref{sqrterror}) individually.
For I, recall that $y_t^* \sim LM(d_y)$ and $\hat{y}_t^* \sim LM(d_{\hat{y}})$. For components II and III, Proposition 1 of \cite{leschinski2017memory} can be used, as it applies to products and squares of long memory time series. Thus, for component II we have
\begin{equation}
\label{prod}
    y_t^* \hat{y}_t^* \sim \left\{\begin{array}{lr}
    LM(d_y + d_{\hat{y}}-1/2), & \text{if}\ S_{y\hat{y}} = 0 \\
    LM(\text{max}\{d_y + d_{\hat{y}} - 1/2, 0\}), & \text{otherwise}, \\
    \end{array}\right.
\end{equation}
where $S_{y\hat{y}} = \sum_{k=-\infty}^{\infty} \gamma_y(k)\gamma_{\hat{y}}(k)$ and $\gamma(k)$ is the autocovariance function.
For component III we get
\begin{equation}
\label{square}
    a_t^{*^2} \sim LM(\text{max}\{2d_a-1/2,0\})\ \text{for}\ a_t^* \in \{y_t^*, \hat{y}_t^*\}.
\end{equation}
Based on $0\leq d_a < 1/2$ for $a_t \in \{y_t, \hat{y}_t\}$, we can generally state that
\begin{equation}
\label{Gen1}
    d_y > d_y + d_{\hat{y}} - 1/2,\ \ d_{\hat{y}} > d_y + d_{\hat{y}} - 1/2
\end{equation}
and
\begin{equation}
\label{Gen2}
    d_{a} > 2 d_a - 1/2.
\end{equation}
\\
The following case-by-case analysis for the expected values derives the memory order of the squared forecast error in Eq. (\ref{sqrterror}).
First, if $\mu_y \neq \mu_{\hat{y}}$, the order of the original terms dominates due to Eq. (\ref{Gen1}) and (\ref{Gen2}), and we get
\begin{equation*}
    d_L = \text{max}\{d_y, d_{\hat{y}}\}.
\end{equation*}
Second, if $\mu_y = \mu_{\hat{y}}$, component I disappears and the linear combination of components II and III remains. It follows from Eq. (\ref{square}) that the lower bound for the memory order of a squared long memory time series is zero and therefore, according to Proposition 3 of \cite{chambers1998long}, the lower bound for the remaining compound term is zero. In addition, the memory order of one of the squared series in Eq. (\ref{square}) always dominates the memory of the product series in Eq. (\ref{prod})
and therefore,
    \begin{equation*}
        d_L = \text{max}\{2 \text{max}\{d_y,d_{\hat{y}}\} - 1/2,0 \}.
    \end{equation*}
\end{proof}

\begin{proof}[\textbf{Proof of Proposition \ref{propFCI}}]
Under the mean relations $\mu_y = \beta_y + \kappa_y\mu_x\ \text{and}\ \mu_{\hat{y}} = \beta_{\hat{y}} + \kappa_{\hat{y}}\mu_x$, Assumption \ref{assumpFCI} and the centered series $x_t^* = x_t - \mu_x$, a similar expression as in Eq. (\ref{sqrterror}) can be obtained.
\begin{align}
\begin{split}
L_t &= (y_t - \hat{y}_t)^2 \\ 
&= (\beta_y + x_t\kappa_y + \epsilon_y)^2 + (\beta_{\hat{y}} + x_t\kappa_{\hat{y}} + \epsilon_{\hat{y}})^2 - 2(\beta_y + x_t\kappa_y + \epsilon_y) (\beta_{\hat{y}} + x_t\kappa_{\hat{y}} + \epsilon_{\hat{y}}) \\
&= (\mu_y + x^*_t\kappa_y + \epsilon_y)^2 + (\mu_{\hat{y}} + x^*_t\kappa_{\hat{y}} + \epsilon_{\hat{y}})^2 - 2(\mu_y + x^*_t\kappa_y + \epsilon_y) (\mu_{\hat{y}} + x^*_t\kappa_{\hat{y}} + \epsilon_{\hat{y}}) \\
&= x_t^{*^2}\kappa_y^2 + 2x_t^*\mu_y\kappa_y + 2x_t^*\kappa_y\epsilon_y + x_t^{*^2}\kappa_{\hat{y}}^2 + 2x_t^*\mu_{\hat{y}}\kappa_{\hat{y}} + 2x_t^*\kappa_{\hat{y}}\epsilon_{\hat{y}}\\ & -2\left[x_t^*\mu_y\kappa_{\hat{y}} + x_t^*\kappa_y\mu_{\hat{y}}+x_t^{*^2}\kappa_y\kappa_{\hat{y}}+x_t^*\kappa_y\epsilon_{\hat{y}} + x_t^*\kappa_{\hat{y}}\epsilon_y \right] \\
& + \epsilon_y^2 + \epsilon_{\hat{y}}^2 + 2\mu_y\epsilon_y + 2\mu_{\hat{y}}\epsilon_{\hat{y}} -2\left[\mu_y\epsilon_{\hat{y}} + \mu_{\hat{y}}\epsilon_y + \epsilon_y\epsilon_{\hat{y}} \right] + \underbrace{\mu_y^2 + \mu_{\hat{y}}^2 - 2\mu_y\mu_{\hat{y}}}_{\text{const.}} \\
& = 2 \Bigl\{ x_t^* [ \kappa_y (\mu_y - \mu_{\hat{y}}) - \kappa_{\hat{y}} (\mu_y - \mu_{\hat{y}}) ] + x_t^* \{ (\kappa_y - \kappa_{\hat{y}})\epsilon_y - (\kappa_y - \kappa_{\hat{y}}) \epsilon_{\hat{y}} \} \\
& + x_t^{*^2} \{ \kappa_y\kappa_{\hat{y}} + 0.5(\kappa_y^2 - \kappa_{\hat{y}}^2) \} + (\mu_y - \mu_{\hat{y}})\epsilon_y - (\mu_y - \mu_{\hat{y}}) \epsilon_{\hat{y}} - \epsilon_y\epsilon_{\hat{y}} \Bigr\} \\
& + \epsilon_y^2 + \epsilon_{\hat{y}}^2 + \text{const.}
\end{split}
\label{sqrterrorFCI}
\end{align}

To determine the memory order of the squared forecast error under fractional cointegration, first note that $x_t^* [ \kappa_y (\mu_y - \mu_{\hat{y}}) - \kappa_{\hat{y}} (\mu_y - \mu_{\hat{y}}) ]$ in Eq. (\ref{sqrterrorFCI}) has a similar structure as $y_t^*(\mu_y-\mu_{\hat{y}}) - \hat{y}_t^*(\mu_y-\mu_{\hat{y}})$ in Eq. (\ref{sqrterror}). The remaining terms in Eq. (\ref{sqrterrorFCI}) are products or squares of centered series and therefore their memory is reduced according to Eq. (\ref{Gen1}) and (\ref{Gen2}). It follows from Assumption \ref{assumpFCI} that $x_t^*$ has the same memory order as the series that are fractionally cointegrated ($y_t$ and $\hat{y}_t$) and also that $d_x > d_{\epsilon_y}, d_{\epsilon_{\hat{y}}}$. Consequently, the reduced memory order of all product and square series is always dominated by the memory of $x_t^*$. Finally, the memory order of the linear combination in Eq. (\ref{sqrterrorFCI}) cannot be less than that of $x_t^*$ whenever a bias term is non-zero. This describes the case of a biased forecast with $\kappa_y \neq \kappa_{\hat{y}}$.\\
To analyze the case of $\kappa_y = \kappa_{\hat{y}}$ with a biased forecast, the term in square brackets in Eq. (\ref{sqrterrorFCI}) is reformulated as $\mu_y(\kappa_y - \kappa_{\hat{y}}) - \mu_{\hat{y}}(\kappa_y - \kappa_{\hat{y}}) = 0$ so that Eq. (\ref{sqrterrorFCI}) simplifies to
\[
    L_t = 2 \{x_t^{*^2} \kappa_y\kappa_{\hat{y}} + (\mu_y - \mu_{\hat{y}})\epsilon_y - (\mu_y - \mu_{\hat{y}}) \epsilon_{\hat{y}} - \epsilon_y\epsilon_{\hat{y}} \} + \epsilon_y^2 + \epsilon_{\hat{y}}^2 + \text{const.}
\]
We know from Eq. (\ref{square}) and (\ref{Gen2}) and from Assumption \ref{assumpFCI} that the memory order is reduced compared to the order of $x_t^*$, but it is not possible to derive the exact memory orders of $\epsilon_y$ and $\epsilon_{\hat{y}}$, and thus we cannot determine whether the memory order of $x_t^{*{^2}}$ or the errors dominate.\\
Finally, we consider the case of an unbiased forecast with $\kappa_y \neq \kappa_{\hat{y}}$. Eq. (\ref{sqrterrorFCI}) then simplifies to
\begin{equation}
\label{balFCIunFOR}
    L_t = \underbrace{2 x_t^{*{^2}}\kappa_y \kappa_{\hat{y}}}_{\text{I}} - \underbrace{2 \epsilon_y\epsilon_{\hat{y}}}_{\text{II}} + \underbrace{\epsilon_y^2 + \epsilon_{\hat{y}}^2}_{\text{III}} +\ \text{const.}
\end{equation}
Again, the memory orders of the individual components in Eq. (\ref{balFCIunFOR}) are determined according to Proposition 1 of \cite{leschinski2017memory}, and finally the memory of the entire series is given by the maximum of the memory orders of the individual components according to \cite{chambers1998long}. We get
\begin{align*}
    \text{I} &\sim LM(\text{max}\{2d_x - 1/2,0\}) \\
    \text{II} &\sim
\left\{
\begin{array}{lr}
    LM(d_{\epsilon_y} + d_{\epsilon_{\hat{y}}} - 1/2),  & \text{if}\ S_{\epsilon_y \epsilon_{\hat{y}}} = 0 \\
    LM(\text{max}\{d_{\epsilon_y} + d_{\epsilon_{\hat{y}}} - 1/2, 0 \}), & \text{otherwise} \\
\end{array}\right. \\
\text{III} &\sim LM(\text{max}\{2d_a - 1/2,0\}),\ \text{for}\ a \in \{\epsilon_y, \epsilon_{\hat{y}} \}.
\end{align*}
Since $d_x > d_{\epsilon_y}, d_{\epsilon_{\hat{y}}}$ holds by Assumption \ref{assumpFCI}, the memory order of component I dominates the memory order of component III and, for the same reason, I also dominates II. Consequently, with an unbiased forecast and $\kappa_y \neq \kappa_{\hat{y}}$, the memory order of the squared forecast error is equal to the order of component I
\begin{equation*}
    d_L = \text{max}\{2d_x - 1/2,0\}.
\end{equation*}
\end{proof}

\begin{proof}[\textbf{Proof of Proposition \ref{Wtest}}]
We follow the proof of \cite{Perr21}, adapting it where necessary. The Wald test for a constant mean against a break at time $T_b(m) = T_b = m_0 + \tau + \floor{\lambda n_0}$ for the loss series $\{L_{t + \tau}^o\}_{t=m}^{T-\tau}$ is

\begin{equation*} \label{eq:test}
    W^m(T_b) = \frac{SSR_{L^o(m)}-SSR(T_b)_{L^o(m)}}{\hat{V}_{L^o(m)}}.
\end{equation*}

For a given $m$, the restricted SSR is

\begin{eqnarray*}
SSR_{L^o(m)} & = & \sum_{t=m}^{T-\tau}L_{t+\tau}^{o^2} - \frac{1}{T-\tau - m+1}\left(\sum_{t=m}^{T-\tau}L_{t+\tau}^{o}\right)^2 \\
& = & \sum_{t=m}^{T-\tau}L_{t+\tau}^{o^2} - \frac{n_0}{T-\tau - m+1}\left(n_0^{-1/2}\sum_{t=m}^{T-\tau}L_{t+\tau}^{o}\right)^2.
\end{eqnarray*}

The unrestricted SSR assuming a break at time $T_b$ is given by

\begin{eqnarray*}
SSR(T_b)_{L^o(m)} & = & \sum_{t=m}^{T_b-\tau}L_{t+\tau}^{o^2} - \frac{n_0}{T_b-\tau - m+1}\left(n_0^{-1/2}\sum_{t=m}^{T_b-\tau}L_{t+\tau}^{o}\right)^2 \\
& + & \sum_{t=T_b-\tau + 1}^{T-\tau}L_{t+\tau}^{o^2} - \frac{n_0}{T-T_b}\left(n_0^{-1/2}\sum_{t=T_b-\tau + 1}^{T-\tau}L_{t+\tau}^{o}\right)^2.
\end{eqnarray*}

We thus have

\begin{eqnarray*}
SSR_{L^{o}(m)} - SSR(T_b)_{L^{o}(m)} & = & -\frac{n_0}{T-\tau - m+1}\left(n_0^{-1/2}\sum_{t=m}^{T-\tau}L_{t+\tau}^{o}\right)^2 \\
& + & \frac{n_0}{T_b-\tau - m+1}\left(n_0^{-1/2}\sum_{t=m}^{T_b-\tau}L_{t+\tau}^{o}\right)^2 \\
& + & \frac{n_0}{T-T_b}\left(n_0^{-1/2}\sum_{t=T_b-\tau + 1}^{T-\tau}L_{t+\tau}^{o}\right)^2.
\end{eqnarray*}

Let $\mu = \lim_{T \rightarrow \infty} \frac{m-m_0}{n_0}$. With $T = n_0 + m_0 + \tau - 1$ and $T_b = m_0 + \tau + \floor{\lambda n_0}$, we have

\begin{align*}
\frac{n_0}{T-\tau -m+1} & = \frac{n_0}{n_0 - (m - m_0)} \rightarrow \frac{1}{1-\mu}, \\
\frac{n_0}{T_b-\tau -m+1} & = \frac{n_0}{\floor{\lambda n_0} - (m - m_0) + 1} \rightarrow \frac{1}{\lambda - \mu}, \\
\frac{n_0}{T - T_b} & = \frac{n_0}{n_0 - \floor{\lambda n_0} - 1} \rightarrow \frac{1}{1 - \lambda}.
\end{align*}

Since $L_{t+\tau -1}^{o} = L_t$ for $t = m+1, \ldots, T - \tau + 1$ and using Assumption \ref{assump}, we get

\begin{eqnarray*}
n_0^{-1/2 + d_L} \sum_{t=m}^{T-\tau}L_{t+\tau}^{o} & = & n_0^{-1/2 + d_L} \sum_{t=m-\tau + 1}^{T-2\tau +1}L_{t+\tau} \\
& = & n_0^{-1/2 + d_L} \sum_{t=m_0 -\tau}^{T-2\tau +1}L_{t+\tau} - n_0^{-1/2 + d_L} \sum_{t=m_0 -\tau}^{m-\tau}L_{t+\tau} \\
& \Rightarrow & \Omega^{1/2}[W_{d_L}(1) - W_{d_L}(\mu)],
\end{eqnarray*}

\begin{eqnarray*}
n_0^{-1/2 + d_L} \sum_{t=m}^{T_b-\tau}L_{t+\tau}^{o} & = & n_0^{-1/2 + d_L} \sum_{t=m-\tau + 1}^{T_b-2\tau +1}L_{t+\tau} \\
& = & n_0^{-1/2 + d_L} \sum_{t=m_0 -\tau}^{T_b-2\tau +1}L_{t+\tau} - n_0^{-1/2 + d_L} \sum_{t=m_0 -\tau}^{m-\tau}L_{t+\tau} \\
& \Rightarrow & \Omega^{1/2}[W_{d_L}(\lambda) - W_{d_L}(\mu)],
\end{eqnarray*}

\begin{eqnarray*}
n_0^{-1/2 + d_L} \sum_{t=T_b - \tau + 1}^{T-\tau}L_{t+\tau}^{o} & = & n_0^{-1/2 + d_L} \sum_{t=T_b -2\tau + 2}^{T-2\tau +1}L_{t+\tau} \\
& = & n_0^{-1/2 + d_L} \sum_{t=m_0 -\tau}^{T-2\tau +1}L_{t+\tau} - n_0^{-1/2 + d_L} \sum_{t=m_0 -\tau}^{T_b-2\tau + 1}L_{t+\tau} \\
& \Rightarrow & \Omega^{1/2}[W_{d_L}(1) - W_{d_L}(\lambda)].
\end{eqnarray*}

Combining these results, we obtain

\begin{multline*}
\hspace{12em}
T^{2d_L}\left(SSR_{L^{o}(m)} - SSR(T_b)_{L^{o}(m)}\right) \Rightarrow \\
\Omega \left[1-\frac{\left[W_{d_L}(1) - W_{d_L}(\mu)\right]^2}{1-\mu} + \frac{\left[W_{d_L}(\lambda) - W_{d_L}(\mu)\right]^2}{\lambda-\mu} + \frac{\left[W_{d_L}(1) - W_{d_L}(\lambda)\right]^2}{1-\lambda}\right],
\hspace{4em}
\end{multline*}

as under the null hypothesis $\hat{V}_{L^{o}(m)} \stackrel{p}{\rightarrow} \Omega$.\\
Note that $\mu = \lim_{T \rightarrow \infty}\frac{m-m_0}{n_0} \leq \lim_{T \rightarrow \infty} \frac{m_1-m_0}{n_0} = \Bar{\mu}$, so that $\mu \in [0, \Bar{\mu}]$. Further, we have for the trimming parameter $\varepsilon$,

\begin{eqnarray*}
T_b & \in & \left[m+\tau+\varepsilon n, m+\tau+(1 - \varepsilon) n\right], \\
\frac{T_b - m_0 - \tau}{n_0} & \in & \left[\frac{(m-m_0)+\varepsilon n}{n_0}, \frac{(m - m_0)+(1-\varepsilon)n}{n_0} \right], \\
\frac{T_b - m_0 - \tau}{n_0} & \in & \left[\frac{(m-m_0)+\varepsilon (n_0 + m_0 - m)}{n_0}, \frac{(m - m_0)+(1-\varepsilon)(n_0 + m_0 - m)}{n_0} \right].
\end{eqnarray*}

Taking the limit implies $\lambda \in [\mu + \varepsilon(1-\mu), 1-\varepsilon(1-\mu)]$, and the result follows.

\end{proof}

\begin{table}[H]
\setlength\tabcolsep{0pt}
\begin{tabular*}{\linewidth}{@{\extracolsep{\fill}} @{}llcccccccc@{} }
\toprule
     & \multicolumn{4}{c}{$\phi=0.1$}              & & \multicolumn{4}{c}{$\phi=0.2$} \\ \cmidrule{2-5} \cmidrule{7-10}
$\alpha/d$ & \multicolumn{1}{c}{0.1} & 0.2 & 0.3 & 0.4 & & 0.1   & 0.2   & 0.3   & 0.4       \\ \cmidrule{1-5} \cmidrule{7-10}
10\% & 7.530 & 7.562 & 7.931 & 8.130 & & 7.327 & 7.611 & 7.937 & 8.246     \\
5\%  & 9.147 & 9.280 & 9.875 & 10.058 & & 8.970 & 9.287 & 9.916 & 10.528    \\
1\%  & 12.984 & 13.000 & 13.943 & 15.867 & & 12.859 & 12.918 & 14.927 & 16.898 \\ \cmidrule{1-5} \cmidrule{7-10}
     & \multicolumn{4}{c}{$\phi=0.3$}              & & \multicolumn{4}{c}{$\phi=0.4$} \\ \cmidrule{2-5} \cmidrule{7-10}
10\% & 7.458 & 7.670 & 8.032 & 8.397 & & 7.494 & 7.570 & 7.719 & 8.413     \\
5\%  & 8.964 & 9.118 & 9.771 & 10.483 & & 9.032 & 9.216 & 9.482 & 10.339    \\
1\%  & 12.402 & 12.885 & 13.805 & 19.454 & & 12.897 & 13.330 & 13.806 & 15.175 \\ \bottomrule
\end{tabular*}
\caption{Critical values for ARFIMA($\phi$,$d$,$0$) processes with $5{,}000$ repetitions, $T=1{,}000$, $m_0 = \floor{0.2T}$, $\Bar{\mu}=0.3$, $\tau=1$ and $\varepsilon=0.1$.}
\label{critAR}
\end{table}

\begin{figure}
    \centering
    \includegraphics[width=1\textwidth]{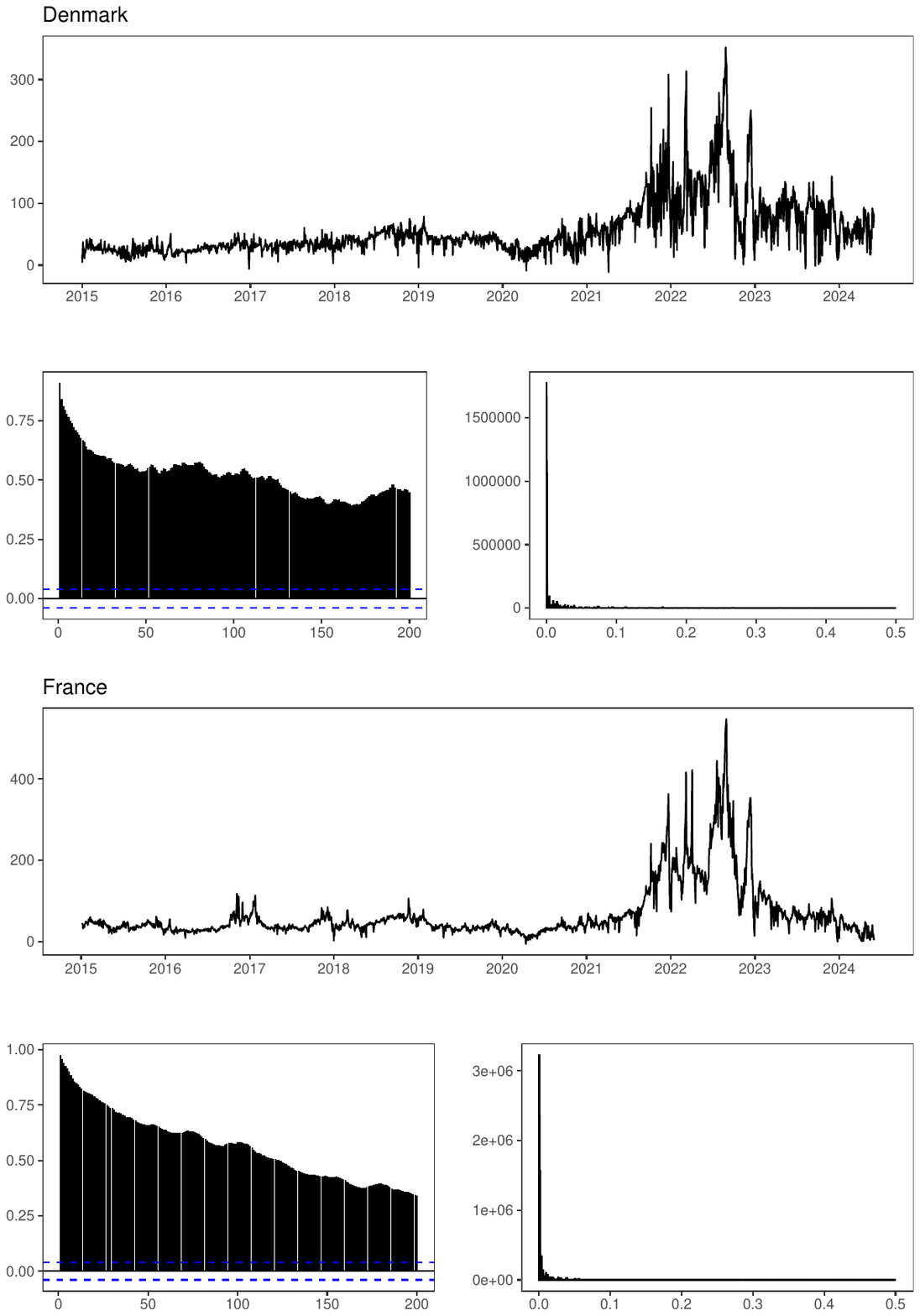}
    \vspace{-50pt}
    \caption{Price series, ACF and periodogram for Denmark and France.}
    \label{DenFra}
\end{figure}

\begin{figure}
    \centering
    \includegraphics[width=1\textwidth]{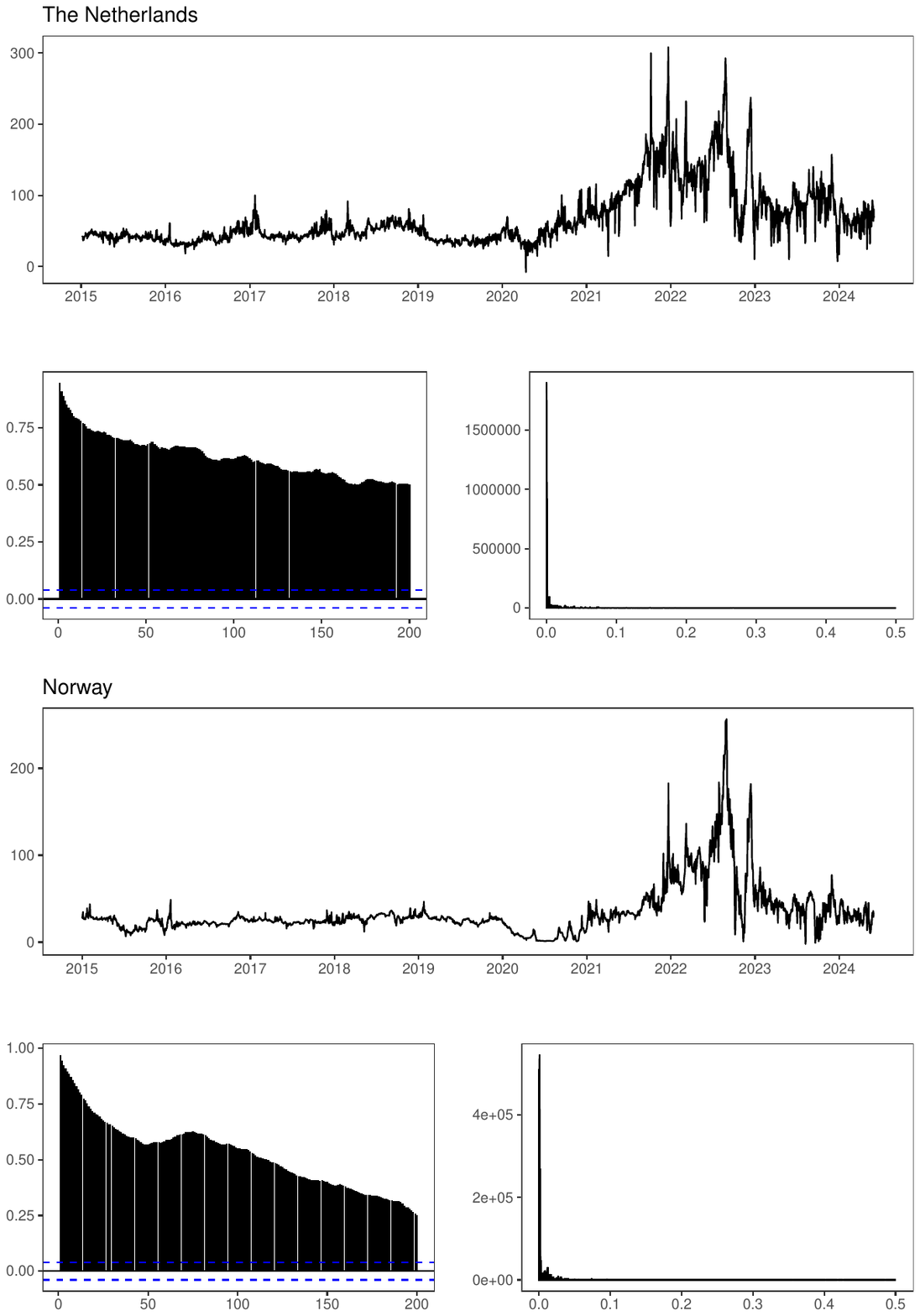}
    \vspace{-50pt}
    \caption{Price series, ACF and periodogram for The Netherlands and Norway.}
    \label{NetNor}
\end{figure}

\begin{figure}
    \centering
    \includegraphics[width=1\textwidth]{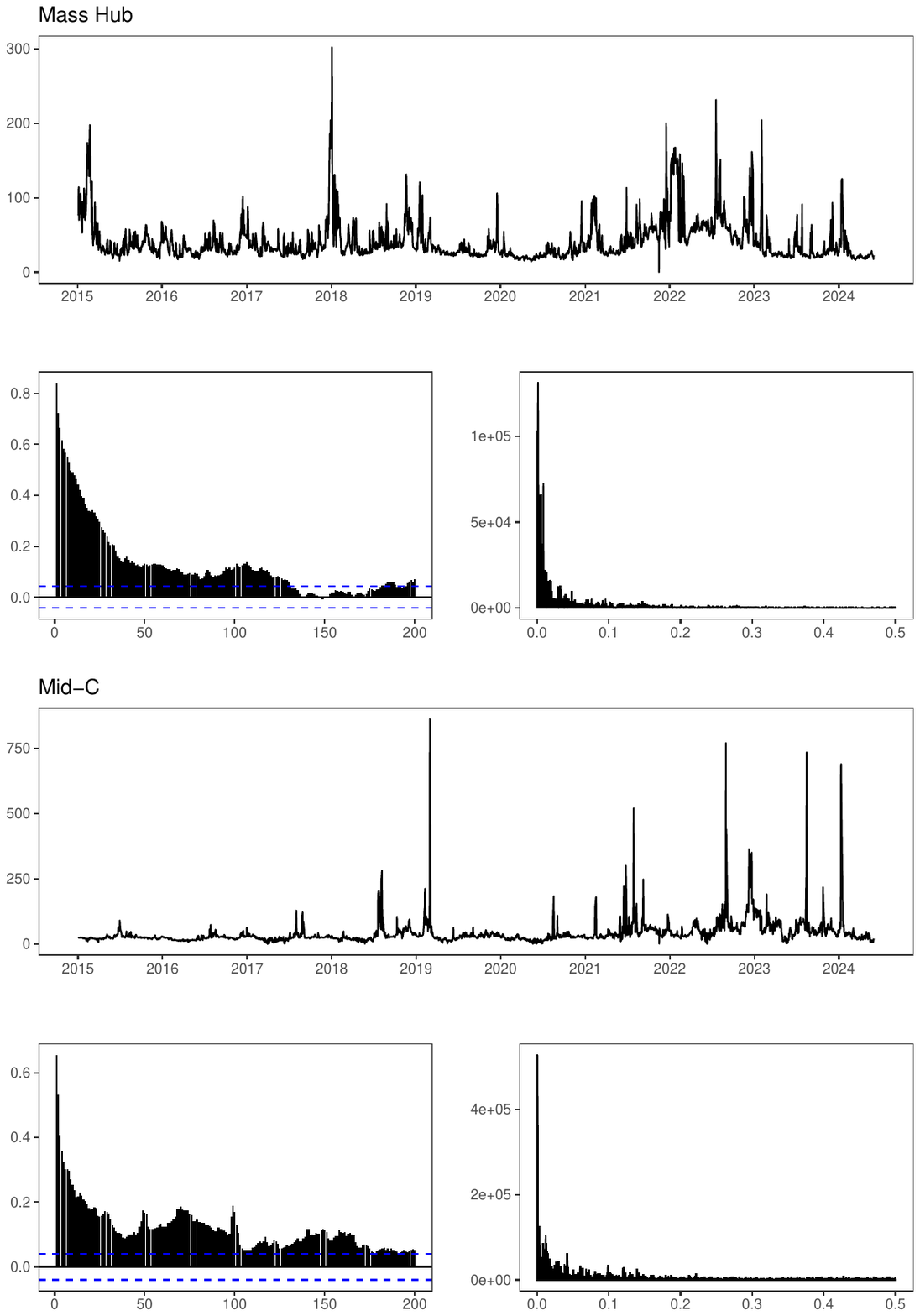}
    \vspace{-50pt}
    \caption{Price series, ACF and periodogram for Mass Hub and Mid-C.}
    \label{MasMid}
\end{figure}

\begin{figure}
    \centering
    \includegraphics[width=1\textwidth]{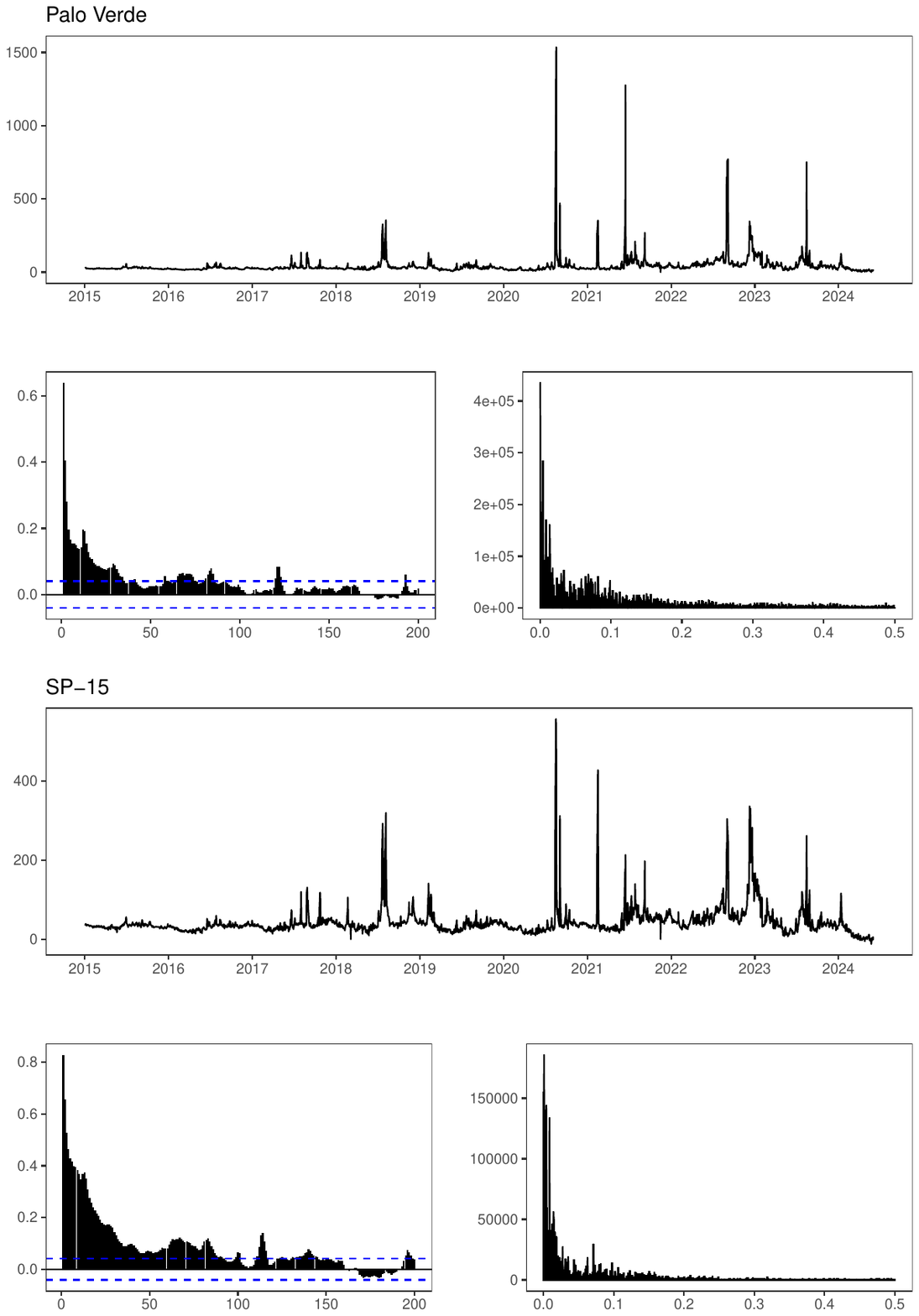}
    \vspace{-50pt}
    \caption{Price series, ACF and periodogram for Palo Verde and SP-15.}
    \label{PalSp15}
\end{figure}


\newpage




\bibliography{biblio}

\begin{thebibliography}{}

\bibitem[\protect\citeauthoryear{Abadir, Distaso, and Giraitis}{Abadir et~al.}{2009}]{abadir2009two}
Abadir, K.~M., W.~Distaso, and L.~Giraitis (2009).
\newblock Two estimators of the long-run variance: beyond short memory.
\newblock {\em Journal of Econometrics\/}~{\em 150\/}(1), 56--70.

\bibitem[\protect\citeauthoryear{Andrews}{Andrews}{1991}]{andrews1991heteroskedasticity}
Andrews, D.~W. (1991).
\newblock Heteroskedasticity and autocorrelation consistent covariance matrix estimation.
\newblock {\em Econometrica: Journal of the Econometric Society\/}, 817--858.

\bibitem[\protect\citeauthoryear{Bai and Perron}{Bai and Perron}{1998}]{bai1998estimating}
Bai, J. and P.~Perron (1998).
\newblock Estimating and testing linear models with multiple structural changes.
\newblock {\em Econometrica\/}, 47--78.

\bibitem[\protect\citeauthoryear{Chambers}{Chambers}{1998}]{chambers1998long}
Chambers, M.~J. (1998).
\newblock Long memory and aggregation in macroeconomic time series.
\newblock {\em International Economic Review\/}, 1053--1072.

\bibitem[\protect\citeauthoryear{Christoffersen and Diebold}{Christoffersen and Diebold}{1997}]{christoffersen1997optimal}
Christoffersen, P.~F. and F.~X. Diebold (1997).
\newblock Optimal prediction under asymmetric loss.
\newblock {\em Econometric theory\/}~{\em 13\/}(6), 808--817.

\bibitem[\protect\citeauthoryear{Diebold and Inoue}{Diebold and Inoue}{2001}]{diebold2001long}
Diebold, F.~X. and A.~Inoue (2001).
\newblock Long memory and regime switching.
\newblock {\em Journal of econometrics\/}~{\em 105\/}(1), 131--159.

\bibitem[\protect\citeauthoryear{Dittmann and Granger}{Dittmann and Granger}{2002}]{dittmann2002properties}
Dittmann, I. and C.~W. Granger (2002).
\newblock Properties of nonlinear transformations of fractionally integrated processes.
\newblock {\em Journal of Econometrics\/}~{\em 110\/}(2), 113--133.

\bibitem[\protect\citeauthoryear{Giacomini and Rossi}{Giacomini and Rossi}{2009}]{giacomini2009detecting}
Giacomini, R. and B.~Rossi (2009).
\newblock Detecting and predicting forecast breakdowns.
\newblock {\em The Review of Economic Studies\/}~{\em 76\/}(2), 669--705.

\bibitem[\protect\citeauthoryear{Granger and Hyung}{Granger and Hyung}{2004}]{granger2004occasional}
Granger, C.~W. and N.~Hyung (2004).
\newblock Occasional structural breaks and long memory with an application to the s\&p 500 absolute stock returns.
\newblock {\em Journal of empirical finance\/}~{\em 11\/}(3), 399--421.

\bibitem[\protect\citeauthoryear{Haldrup and Nielsen}{Haldrup and Nielsen}{2006}]{haldrup2006regime}
Haldrup, N. and M.~{\O}. Nielsen (2006).
\newblock A regime switching long memory model for electricity prices.
\newblock {\em Journal of econometrics\/}~{\em 135\/}(1-2), 349--376.

\bibitem[\protect\citeauthoryear{Horv{\'a}th and Kokoszka}{Horv{\'a}th and Kokoszka}{1997}]{horvath1997effect}
Horv{\'a}th, L. and P.~Kokoszka (1997).
\newblock The effect of long-range dependence on change-point estimators.
\newblock {\em Journal of Statistical Planning and Inference\/}~{\em 64\/}(1), 57--81.

\bibitem[\protect\citeauthoryear{Kruse, Leschinski, and Will}{Kruse et~al.}{2018}]{Krus18}
Kruse, R., C.~Leschinski, and M.~Will (2018).
\newblock Comparing predictive accuracy under long memory, with an application to volatility forecasting.
\newblock {\em Journal of Financial Econometrics\/}~{\em 17\/}(2), 180--228.

\bibitem[\protect\citeauthoryear{K{\"u}nsch}{K{\"u}nsch}{1987}]{kunsch1987statistical}
K{\"u}nsch, H.-R. (1987).
\newblock Statistical aspects of self-similar processes.
\newblock {\em Proceedings of the First World Congress of the Bernoulli Society\/}~{\em 1}, 67--74.

\bibitem[\protect\citeauthoryear{Leschinski}{Leschinski}{2017}]{leschinski2017memory}
Leschinski, C. (2017).
\newblock On the memory of products of long range dependent time series.
\newblock {\em Economics Letters\/}~{\em 153}, 72--76.

\bibitem[\protect\citeauthoryear{Mikosch and St{\u{a}}ric{\u{a}}}{Mikosch and St{\u{a}}ric{\u{a}}}{2004}]{mikosch2004nonstationarities}
Mikosch, T. and C.~St{\u{a}}ric{\u{a}} (2004).
\newblock Nonstationarities in financial time series, the long-range dependence, and the igarch effects.
\newblock {\em Review of Economics and Statistics\/}~{\em 86\/}(1), 378--390.

\bibitem[\protect\citeauthoryear{Paza~Mboya and Sibbertsen}{Paza~Mboya and Sibbertsen}{2023}]{paza2023optimal}
Paza~Mboya, M. and P.~Sibbertsen (2023).
\newblock Optimal forecasts in the presence of discrete structural breaks under long memory.
\newblock {\em Journal of Forecasting\/}~{\em 42\/}(7), 1889--1908.

\bibitem[\protect\citeauthoryear{Perron}{Perron}{2006}]{perron2006dealing}
Perron, P. (2006).
\newblock Dealing with structural breaks.
\newblock {\em Palgrave handbook of econometrics\/}~{\em 1\/}(2), 278--352.

\bibitem[\protect\citeauthoryear{Perron and Yamamoto}{Perron and Yamamoto}{2021}]{Perr21}
Perron, P. and Y.~Yamamoto (2021).
\newblock Testing for changes in forecasting performance.
\newblock {\em Journal of Business \& Economic Statistics\/}~{\em 39\/}(1), 148--165.

\bibitem[\protect\citeauthoryear{Robinson}{Robinson}{1995}]{robinson1995gaussian}
Robinson, P.~M. (1995).
\newblock Gaussian semiparametric estimation of long range dependence.
\newblock {\em The Annals of statistics\/}, 1630--1661.

\bibitem[\protect\citeauthoryear{Robinson}{Robinson}{2005}]{robinson2005robust}
Robinson, P.~M. (2005).
\newblock Robust covariance matrix estimation: Hac estimates with long memory/antipersistence correction.
\newblock {\em Econometric Theory\/}~{\em 21\/}(1), 171--180.

\bibitem[\protect\citeauthoryear{Rossi and Inoue}{Rossi and Inoue}{2012}]{rossi2012out}
Rossi, B. and A.~Inoue (2012).
\newblock Out-of-sample forecast tests robust to the choice of window size.
\newblock {\em Journal of Business \& Economic Statistics\/}~{\em 30\/}(3), 432--453.

\bibitem[\protect\citeauthoryear{Wang}{Wang}{2008}]{wang2008change}
Wang, L. (2008).
\newblock Change-in-mean problem for long memory time series models with applications.
\newblock {\em Journal of Statistical Computation and Simulation\/}~{\em 78\/}(7), 653--668.

\end{thebibliography}
\end{document}